\documentclass{article}
\usepackage[utf8]{inputenc} 
\usepackage{amsfonts}
\usepackage{dsfont}
\let\mathbb\mathds 
\usepackage{mathtools} 
\usepackage{xcolor} 
\usepackage{soul}
\usepackage{verbatim}
\usepackage{authblk}

\usepackage[colorlinks=true,linkcolor=orange,urlcolor=orange,citecolor=blue]{hyperref} 
\usepackage{float}
\usepackage[bottom]{footmisc} 
\usepackage[labelsep=period, labelfont=bf]{caption} 
\usepackage{subcaption} 
\usepackage{float,subfloat} 
\usepackage{physics,derivative} 
\usepackage{accents} 
\usepackage{scalefnt} 
\usepackage{alphabeta} 
\usepackage{relsize}

\usepackage{booktabs}
\usepackage{graphicx} 
\usepackage{multicol,multirow} 

\usepackage{adjustbox} 
\usepackage{pdfpages}

\usepackage{yhmath} 
\usepackage{eurosym} 

\usepackage{geometry}
\usepackage{setspace}
\usepackage[skip=5pt plus1pt,indent=2em]{parskip}

\usepackage[sort]{natbib} 
\title{Treatment effects for marginal decision-makers:\\ Everyone is marginal}  
\author[]{Haotian Deng}

\affil[]{Department of Economics, Ghent University\footnote{
    This is a preliminary version of the my thoughts as a note and is thus not in a paper state. You can contact me at \href{mailto:haotian.deng@ugent.be}{haotian.deng@ugent.be} for a discussion.  }
}

\date{29 August 2025}

\begin{document} 

\maketitle 

\begin{abstract}

    This paper develops a framework for identifying treatment effects when a policy simultaneously alters both the incentive to participate and the outcome of interest---such as hiring decisions and wages in response to employment subsidies; or working decisions and wages in response to job trainings. This framework was inspired by my PhD project on a Belgian reform that subsidised first-time hiring, inducing entry by marginal firms yet meanwhile changing the wages they pay. Standard methods addressing selection-into-treatment concepts (like Heckman selection equations and local average treatment effects), or before-after comparisons (including simple DiD or RDD), cannot isolate effects at this shifting margin where treatment defines who is observed. I introduce marginality-weighted estimands that recover causal effects among policy-induced entrants, offering a policy-relevant alternative in settings with endogenous selection. This method can thus be applied widely to understanding the economic impacts of public programmes, especially in fields largely relying on reduced-form causal inference estimation (e.g. labour economics, development economics, health economics).  (\emph{JEL codes:} C01, C10, J01, J08) \\


\end{abstract}

\section{Introduction} 

This note develops a theoretical framework for identifying treatment effects when the treatment itself arises from an endogenous, policy-sensitive decision, and then applies this framework to the empirical question of payroll tax incidence in small firms for their initial hiring. In this setting, outcomes (wages) are only defined for agents who enter the participation (to become an employer), yet the decision to participate is precisely what the policy intends to influence. This endogeneity creates not only an empirical but a fundamentally conceptual challenge: treatment status is not merely a covariate but a defining threshold that alters who is observed and whether the outcome of interest is defined at all. We thus have our question regarding \textbf{identification:} what should researchers identify?

Economists have long recognised the problem of selection into treatment. The local average treatment effect (LATE) framework \citep{imbens1994identification} and its extensions, including marginal treatment effects (MTE) \citep{heckman2005structural}, formalized causal inference under selection, especially in the presence of continuous unobserved heterogeneity. These approaches remain influential, but they are now decades old and rooted in a reduced-form tradition. They do not distinguish between entry into treatment and variation within treatment, and they treat marginal and inframarginal participants as observationally similar except for an index of resistance to treatment.

This note takes a different stance. I focus on settings where entry into treatment is itself an economic outcome, and where agents who enter and those who do not are structurally different. The empirical motivation of this note is a 2016 Belgian payroll tax reform that granted a permanent exemption to firms that hired their first employee. The policy targeted non-employers at the extensive margin, aiming to induce new firms' labour market participation. The treatment of a hiring subsidy coincides with participation of the act of hiring, and therefore defines the sample of observed outcomes and changes the composition of the employer population. 

The realistic goal is to study the reform effects on entrant employers' wage (rates) offered, and then answer the payroll tax incidence for the first hires. To do so, I first develop a model of heterogeneous firms, where each firm's productivity determines both its hiring probability and the wage it would pay if it does hire. The reform shifts the hiring probability function, inducing entry by marginal firms that would not have hired otherwise. I then show that observed differences in post- and pre-reform average outcomes, typically yielded by most methods, can be decomposed into causal treatment effects plus (uninterpretable) biases. 

To fix these problems, I introduce a class of importance-weighted estimands that target the policy-induced margin, and relate them to commonly applied estimation techniques. 
This framework thus bridges structural and reduced-form perspectives and provides a policy-relevant estimand when both treatment take-up and outcome definitions are endogenous. It is not a reinterpretation of LATE or MTE, but an attempt to address the empirical challenges that arise when treatment status defines existence itself. 

My setting parallels that studied by \citet{lee2009training}, who addresses a similar `double selection' (on both the existence and the outcome) structure in evaluating the effects of job training programs. Lee derives sharp nonparametric bounds on treatment effects under minimal assumptions by trimming treated observations to match the support of the control group, a strategy that ensures comparability even when potential outcomes are partially missing.

Instead of bounding, the approach developed in this paper explicitly models the selection process through a productivity-indexed hiring probability function. This allows for the construction of meaningful estimands under the shifting employer distribution, including importance-weighted and marginality-weighted treatment effects. These estimands remain interpretable even when the support of treated and untreated groups does not fully overlap, and they offer a more informative alternative in settings where the policy causes substantial endogenous entry into the treated population.

\section{Model of programme participation} 

This section builds a minimal note summarising the entrepreneurial decision model developed by \citet{deng2025paper0}, a microeconomic-theory framework of employer firm formation regarding whether entrepreneurs choose to become an employer or stay as solo self-employed. Although the model wording literally addresses entrepreneurship, it can be easily extended to general programme participation situations. 

A continuum of small entrepreneurial firms differ in their (unobserved) productivity $\theta_i \sim f(.)$ over $\mathbb{R}^+$, which is the population density; the cumulative distribution density is $F(\theta)=\int_{0}^{\theta}{f(x)dx}$. Firms with a higher productivity level can, with production factor inputs constant, generate higher production. They enter the labour market by hiring their first employees with a probability $p^s(\theta_i)$ and if they do, they offer a wage rate $y^s(\theta_i)$, differently in an untreated status (no such exemption, $s=0$, untreated outcome $y^0$) and in a treated status (with such exemption, $s=1$, trea`ted outcome $y^1$). The rest stays self-employed with no employees and do not pay wages. 

With minimal standard assumptions, the model concludes that (1) the probability of hiring is an increasing and continuous function in the productivity level, and (2) the reform increases the probability of entry from $p^0(\theta_i)$ to $p^1(\theta_i)$ and therefore raises the mass of entrant employers from $N^0=\int_{0}^{+\infty}{p^0(\theta)dF(\theta)}$ to $N^1=\int_{0}^{+\infty}{p^1(\theta)dF(\theta)}$.

An important consequence of the hiring probability is that firms of different productivity levels occurs in the employer dataset with different probabilities. The probabilistic distribution of employer firms in state $s$ is a importance-weighted distribution, a composite of $p^s(\theta)$ and $f(\theta)$, 
\begin{equation}
    \label{eq:Q} 
    Q^s(\theta') = \Pr\{\theta \le \theta' | \text{state }s \} = \int_{0}^{\theta'}{q^s(x)dx},  
\end{equation}
with $q^s(\theta)=\frac{p^s(\theta)}{N^s}f(\theta)$ the importance-weighted probability density. Under this density, the observed mean wage (or other outcomes) is 
\begin{equation}
\label{eq:observed.mean}
    \bar y^s= \mathbb{E}_{Q^s}[y^s(\theta)] = \int_{0}^{+\infty}{y^s(\theta) \frac{p^s(\theta)}{N^s} dF(\theta)} = \mathbb{E}_{F}\bigg[y^s(\theta)\frac{p^s(\theta)}{N^s}\bigg], 
\end{equation}
with importance weights $\frac{p^s(\theta)}{N^s}$.\footnote{This is the importance weight when integrated against $f(\theta)d\theta$. When integrated against $d\theta$ alone, the weights are $\frac{p^s(\theta)f(\theta)}{N^s}$.}  Here I borrow the term of `importance weights' from the field of machine learning \citep{sugiyama2007importance.weighted}, but the idea is simple in economics: the composition of employer entrants changed from distribution $Q^0$ to $Q^1$. There are possibly many other terms that can apply. 

\section{(Possibly) Bad estimands} 

\subsection{Population average treatment effect} 
\label{sec:pate}
If we know the potential outcome dependence on firm types $y^1(.), y^0(.)$, then the treatment effect for type $\theta$ firm is $\tau(\theta) \coloneqq y^1(\theta)-y^0(\theta)$, and the population average treatment effect (PATE) weighted under the productivity distribution scheme $F(\theta)$ is 
\begin{equation}
    \tau^{\texttt{PATE}} = \mathbb{E}_{F}[\tau(\theta)] = \int_{0}^{+\infty}{ \tau(\theta) dF(\theta)}.
\end{equation}
The PATE neglects the probability of hiring and only weights according to the population distribution of types; as a result, it implies that firms of low productivity (which basically never hire) are equally important as firms of high productivity (which always hire) in drawing the average treatment effect. 

While this parameter appears intuitive and is often regarded as a natural benchmark, it suffers from a fundamental shortcoming in the context of employer-based policies: it averages over all firm types, including those that never hire employees and thus never appear in observed wage data. In reality, the average among employer firms always carries some importance weighting: the observed means are additionally weighted with firms' probability of hiring (and thus occurring in the sample). Therefore, PATE is not sufficiently interesting to know. 

Inverse probability weighting (IPW) estimators \citep{rosenbaum1983central, hirano2003efficient, imbens2004nonparametric}, commonly used in applied econometrics, often aim to recover this PATE by reweighting observations to match the population distribution (see Appendix \ref{app:ipw.recovers.pate} for a simple derivation). However, in doing so, they neglect a key feature of the data-generating process: firms enter the employer sample with unequal probability, determined endogenously by their productivity and the policy environment. In practice, observed outcomes (e.g., wages) are only defined for employer firms, and these are not a random draw from the population of all firms. They are selectively observed, based on hiring thresholds that the policy itself shifts. 

Moreover, the reference distribution \( F(\theta) \) itself is not uniquely determined. In practice, the composition of the population depends on whether firms that never hire (purely self-employed entities) are included. This choice affects the weight assigned to firm types that are observationally irrelevant for the outcome of interest. If all legal entities are included, \( F(\theta) \) will place substantial mass on types that contribute no information on wages. If only firms with positive hiring probability are included, then the estimand already deviates from a true population average. 

\subsection{Observed mean difference estimand}
Designing an identification strategy for the treatment effect on wages is complicated by the fact that wages are only defined for employer firms, and the set of employers itself changes endogenously in response to the reform. Before the reform, firms that would later become treated are non-employers, and thus do not generate wage observations. As a result, some standard estimations like a panel difference-in-differences approach are infeasible: there is no pre-treatment wage trajectory for firms that were not yet hiring.

To leverage the policy variation over time, a natural alternative is to compare post-reform cohorts of new employers to pre-reform cohorts, and to adopt a cross-sectional difference-in-differences design or a regression-discontinuity-in-time design by leveraging the cohort time. Assuming for now that a valid control group has been constructed to account for concurrent trends (in a DiD) or that outcome polynomials are correctly modelled as functions of the entry time (in a RDiT), such a strategy yields the observed mean difference (OMD) estimand:
\begin{equation}
\label{eq:def.omd}
    \tau^\texttt{OMD} = \bar y^1 - \bar y^0 
    , 
\end{equation}
with the observed means naturally carrying importance weighting schemes as defined in \eqref{eq:observed.mean}, differently in the pre-reform periods and in the post-reform periods. 

Although feasible to calculate, the OMD estimand is uninteresting as it carries two serious biases along with an interpretable treatment effect parameter. It can be decomposed into 
\begin{equation}
\label{eq:omd.de}
    \tau^\texttt{OMD} = \underbrace{\mathbb{E}_{Q^1}[\tau(\theta)]}_\text{Post-reform-weighted ATE}
    + \underbrace{\mathbb{E}_{Q^1}[y^0(\theta)]-\mathbb{E}_{Q^0}[y^0(\theta)]}_\text{Selection bias} 
    + \underbrace{\mathbb{E}_{F}\bigg[ y^0(\theta)p^0(\theta)\bigg(\frac{1}{N^1} - \frac{1}{N^0}\bigg) \bigg]}_\text{Re-weighting effect bias}, 
\end{equation}
where: the first term is a true ATE weighted under the post-reform hiring probability, an interpretable parameter; the second term is a selection bias equal to the difference in the untreated wages between two different distributions, which arises from the composition changes in firm productivity levels among employer firms; the third term is a re-weighting effect bias equal to the difference in untreated wages between two weighting schemes, which occurs solely because the relative importance of firms in the hiring distribution changes following the reform. 

In this Eq \eqref{eq:omd.de}, the first term $\mathbb{E}_{Q^1}[\tau(\theta)]$ weights the average treatment effects according to the post-reform importance-weighted distribution, giving more weights to firms that are more likely to hire, and fewer weights to firms that are less likely. This weighting scheme is consistent with observation in reality and is thus interesting to know. Alternatively, we can write the first term evaluated under the pre-reform importance-weighted distribution, $\mathbb{E}_{Q^0}[\tau(\theta)]$, plus the two alternatively defined biases (See Appendix \ref{app:alt.omd}). 

\subsection{Problems with the marginal--inframarginal dichotomy}

A third common but problematic approach is to interpret the reform through a dichotomy of marginal versus inframarginal firms. The intuition is that some firms are induced by the policy to cross the threshold into treatment (`marginals'), while others would have hired regardless (`inframarginals'). This classification may appear appealing and useful in many applications (e.g., \citealp{hombert2020france, branstetter2014entry}), but here it relies on imposing an arbitrary threshold on the participation probability.

Formally, suppose one specifies a participation probability threshold \( \underline{p} \) and defines the productivity cut-off points \( \underline{\theta} \) and \( \underline{\underline{\theta}} \) such that
\[
p^0(\underline{\theta}) = \underline{p}, \quad p^1(\underline{\underline{\theta}}) = \underline{p}.
\]
The policy-induced shift in the hiring function from \( p^0 \) to \( p^1 \) ensures that \( \underline{\underline{\theta}} < \underline{\theta} \), and allows one to define two conditional average treatment effects:
\begin{align}
    \tau^\texttt{Infra}(\underline{p}) &\coloneqq \mathbb{E}[\tau(\theta) \mid \theta > \underline{\theta}], \\
    \tau^\texttt{Mar}(\underline{p}) &\coloneqq \mathbb{E}[\tau(\theta) \mid \underline{\underline{\theta}} < \theta \le \underline{\theta}].
\end{align} 

\begin{figure}[h!]
    \centering
    \includegraphics[width=\linewidth]{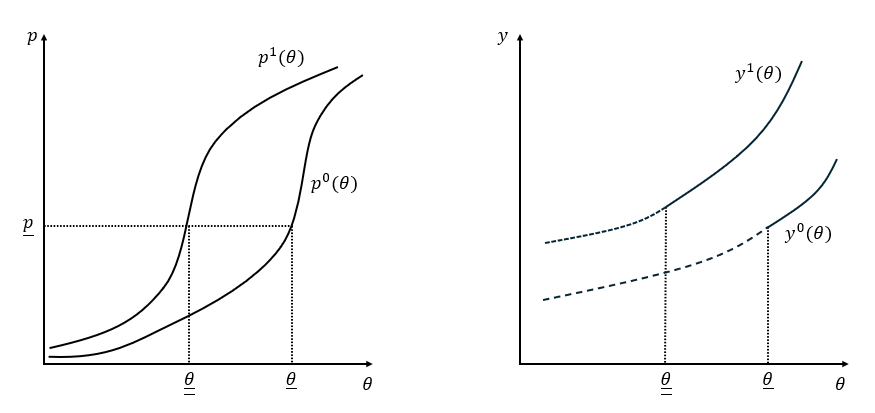}
    \caption{Probability of hiring and potential outcomes in a marginal--inframarginal dichotomy}
    \label{fig:infra.mar}
\end{figure}

However, this construction raises several conceptual problems. First, the threshold \( \underline{p} \) is not identified from the model or data but must be imposed manually. Different choices of \( \underline{p} \) result in different delineations of marginal and inframarginal types, and therefore alter the interpretation of the resulting estimands. This arbitrariness undermines both the stability and the policy relevance of the marginal--inframarginal split.

Second, the imposition of a sharp threshold \( \underline{p} \) discretises an inherently continuous decision process. The hiring probability function \( p(\theta) \) is continuous in firm productivity by construction, and the reform shifts this entire function upward. Categorising firms as either marginal or inframarginal discards this structure and replaces it with a rigid partition that lacks behavioural foundation.

Third, and more seriously, the marginal group defined as \( \theta \in (\underline{\underline{\theta}}, \underline{\theta}] \) consists of firms that do not hire in the pre-reform regime. As such, their untreated outcomes \( y^0(\theta) \) are never observed by construction. This creates a fundamental counterfactual problem: there is no empirical support for their behaviour in the absence of treatment, making \( \tau^\texttt{Mar}(\underline{p}) \) unidentified without strong structural assumptions or extrapolations from other groups. 

To make this concrete, the marginal average treatment effect can be written as:
\begin{equation}
\tau^\texttt{Mar}(\underline{p}) = \mathbb{E}\left[ y^1(\theta) \mid \underline{\underline{\theta}} < \theta \le \underline{\theta} \right] - \mathbb{E}\left[ y^0(\theta) \mid \underline{\underline{\theta}} < \theta \le \underline{\theta} \right].
\end{equation} 
The first term, the post-reform mean among induced entrants, is well-defined and observable. The second term, however, is undefined: these firms were non-employers before the reform and thus paid no wages. Their counterfactual wage outcomes \( y^0(\theta) \) cannot be observed or inferred without unverifiable extrapolation (see the right panel of Figure \ref{fig:infra.mar}). (This is a parallel problem to the ``no common support'' in the propensity score matching estimator methods.) 

To isolate the issue of the dichotomous classification itself, the notation here suppresses any subscript indicating the weighting distribution (e.g., \( Q^0 \), \( Q^1 \)). But in any realistic analysis, where one must average these conditional effects under a coherent probability distribution, the arbitrariness of the dichotomy becomes even more problematic. It is unclear under which regime (pre- or post-reform) these averages should be evaluated, since the marginal group exists only under the post-reform distribution and is absent by construction in the pre-reform counterfactual.

\section{Good estimands}

The issues highlighted in the previous section—particularly the observational detachment of the PATE, the bias-prone nature of the observed mean difference (OMD), and the counterfactual gaps in the marginal-inframarginal framework—call for estimands that are better grounded in observed data, more coherent with treatment-induced selection, and more meaningful for policy analysis. In this section, I define a class of treatment effect estimands that are constructed via continuous reweighting of the population distribution, using the selection structure of the model to determine the relevant subpopulations for averaging.

\subsection{Importance-weighted average treatment effects}
\label{sec:q0q1}

A natural starting point is to define treatment effects not over the entire firm population, but over the subpopulations of firms that actually hire under each regime. This gives rise to two importance-weighted average treatment effects:
\begin{align}
\tau^{Q^0} &\coloneqq \mathbb{E}_{Q^0}[\tau(\theta)] = \int \tau(\theta) \frac{p^0(\theta)}{N^0} dF(\theta), \\
\tau^{Q^1} &\coloneqq \mathbb{E}_{Q^1}[\tau(\theta)] = \int \tau(\theta) \frac{p^1(\theta)}{N^1} dF(\theta).
\end{align}

These estimands address key shortcomings of the bad estimands. Unlike the PATE, they assign realistic weights to firms of different hiring probability (productivity). Unlike the OMD, they apply consistent weights across potential outcomes and thus avoid conflating treatment effects with selection or compositional changes. And unlike the marginal--inframarginal split, they do not rely on arbitrary thresholds or unobserved counterfactuals.

The choice between \( Q^0 \) and \( Q^1 \) depends on the policy question. Weighting under \( Q^0 \) corresponds to the effect on firms that would have hired in the absence of the reform---akin to an average treatment on the treated (ATT) among inframarginals. Weighting under \( Q^1 \) instead focuses on the firms who actually hired after the reform, reflecting the realised composition of treated firms. Both estimands are well-defined and feasible to estimate from observed data, as the sample of employers under each regime provides direct support for each weighting function.

Propensity score matching (PSM) \citep{rosenbaum1983central, heckman1997matching, abadie2006matching}, another commonly applied method (cf. IPW in Section \ref{sec:pate}), can recover these two estimands if performed properly (see Appendix \ref{app:psm} for proofs). Matching firms based on the pre-reform hiring probability scheme \( p^0(.) \), by fitting only pre-reform data and extrapolating to post-reform, recovers a \( Q^0 \)-weighted treatment effect: the average treatment effect among firms that would have hired in the absence of the reform. Similarly, matching based on \( p^1(.) \), by fitting only post-reform data and extrapolating to pre-reform, recovers the \( Q^1 \)-weighted ATE among firms that do hire after the reform. However, matching \( p^0(.) \) to \( p^1(.) \), or pooling observations across regimes to estimate a single propensity score scheme, risks invalid inference (and results in uninterpretable estimands like the OMD): the propensity scores originate from structurally different selection processes and cannot be reconciled. 

Applicability is limited in their own aspects. \( \hat{p}^0 \)-matching might break down for marginal entrants since their pre-reform counterparts did not hire and thus lack observed outcomes. The result is either a loss of support (excluding marginals) or bias due to extrapolation from non-observed counterfactuals. 
Conversely, \( \hat{p}^1 \)-matching faces the issue that pre-reform firms did not face the same incentives as post-reform firms. But since both marginals and inframarginals are captured by \( \hat{p}^1(X) \), this matching should remain more feasible if one wants to include marginals in the estimation.    

\subsection{Generalised weighting schemes}

The two importance-weighted estimands above can be seen as special cases of a broader framework. More generally, one can define:
\begin{equation}
\tau^w \coloneqq \frac{\int \tau(\theta) \cdot w(\theta) \, dF(\theta)}{\int w(\theta) \, dF(\theta)},
\end{equation}
where \( w(\theta) \) is any non-negative, integrable weighting function. This formulation allows analysts to tailor the estimand to the policy objective or empirical setting. For instance: Choosing \( w(\theta) = p^0(\theta) \) recovers \( \tau^{Q^0} \), choosing \( w(\theta) = p^1(\theta) \) recovers \( \tau^{Q^1} \), and choosing \( w(\theta) = 1 \) recovers the (uninformative) PATE.

This generalised framework resolves the rigidity of the marginal--inframarginal split by treating selection as continuous and policy-dependent. Moreover, it avoids the inconsistencies of the OMD by applying symmetric and interpretable weights across treatment and control potential outcomes.

From a feasibility perspective, however, estimating \( \tau^w \) requires knowledge or estimation of the weighting function \( w(\theta) \), and, in general, access to both \( y^1(\theta) \) and \( y^0(\theta) \) across the support of \( w(\theta) \). When \( w \) corresponds to observed hiring probabilities (as in \( Q^0 \) or \( Q^1 \)), this is feasible from the data. For other \( w \), feasibility depends on whether those weights are supported in both regimes or can be estimated structurally.

\subsection{Marginality-weighted treatment effect}
\label{sec:marginality}

A particularly policy-relevant weighting function arises by considering the shift in hiring probability induced by the reform:
\[
\Delta p(\theta) \coloneqq p^1(\theta) - p^0(\theta).
\]
This function captures the incremental participation probability at each productivity level and defines the marginality-weighted average treatment effect:
\begin{equation}
\tau^{\Delta p} \coloneqq \frac{\int \tau(\theta) \cdot \Delta p(\theta) \, dF(\theta)}{\int \Delta p(\theta) \, dF(\theta)}.
\end{equation}

This estimand directly targets the subpopulation of firms whose hiring decision was responsive to the reform. It avoids the thresholding problem of the marginal--inframarginal dichotomy by explicitly modelling marginality as a continuous concept. 

If we can capture good proxies of the unobservable firm productivity into a set of observables $X$, under weak and interpretable conditions, it can be shown that the estimand can be rewritten entirely in terms of observables. Define \( \tau(x) = \mathbb{E}[\tau(\theta) \mid X = x] \) and \( \Delta p(x) = \mathbb{E}[\Delta p(\theta) \mid X = x] \), then the marginality-weighted estimand can be expressed as
\begin{equation} 
\tilde\tau^{\Delta p} = \frac{\int \tau(x) \cdot \Delta p(x) \, dF(x)}{\int \Delta p(x) \, dF(x)}.
\end{equation}
This observable analogue can be consistently estimated by partitioning the support of \( X \), estimating \( \Delta p(x) \) and \( \tau(x) \) within each cell, and applying a weighted average. A formal derivation is given in Appendix \ref{app:delta_p_observable}.

An additional practical advantage of the marginality-weighted estimand is that it relies only on estimating the policy-induced \emph{change} in hiring probability, \( \Delta p(x) = p^1(x) - p^0(x) \), rather than the full levels of the regime-specific hiring propensities \( p^0(x) \) and \( p^1(x) \) themselves. This distinction is important: while the \( Q^0 \)- and \( Q^1 \)-weighted estimands require full propensity models—typically via possibly non-linear methods across the entire support of \( x \), the marginality-weighted estimand treats the difference in probabilities as the estimand of interest. In practice, estimating a linear model in probability is often more robust and less sensitive to misspecification than estimating levels, particularly in regions where the propensity is close to zero or one. This makes \( \tau^{\Delta p} \) not only more policy-relevant but also more empirically feasible in applied work. 


\section{Conclusion}

In conclusion, the framework developed here highlights the central role of marginal decision-makers in causal analysis. Treatment not only shifts outcomes but also determines the very existence of the relevant units, making participation endogenous and outcomes only meaningful for those who cross the margin. This perspective unifies diverse applications under a single principle: \emph{everyone is marginal with respect to some decision}. 

Through a long discussion, the method proposed in the Section \ref{sec:marginality} is my favourite one and is the innovation of this paper due to two important reasons. First, this is actually the most realistically relevant parameter for the programmes that are designed to encourage people at the margin to participate, where policymakers care about the economic effects for these marginal people induced to enter instead of the effects for inframarginal people who enter anyway. Second, as I discussed and shall discuss more, this parameter is easier to recover than those proposed in Section \ref{sec:q0q1}. In a next version of the note, I plan to formally propose estimators for the marginality-weighted estimands and prove their statistical properties. 

Just as an example, I list four cases where my framework can be highly useful for real-world understanding:
\begin{enumerate}
\item 
If a government provides a subsidy to firms that hire their first employee, how should researchers define and measure the treatment effect on wages, given that the existence of the first employee depends on the treatment itself?
\item 
When scholarships are awarded only to students who are admitted to university, how should researchers define the treatment effect on grade point average (GPA), recognising that university grades are only defined for those who do get admitted and enrolled?
\item 
If immigrants receive financial or institutional aid to facilitate their settlement, how should researchers define the treatment effect on an integration index, acknowledging that integration only matters for people who choose to immigrate?
\item 
If advanced spaceships were developed to enable migration to Mars, how should researchers know the treatment effect on life expectancy, given that human life on Mars is only made possible by the spaceships?
\end{enumerate}
Research questions and policies alike are commonplace in the fields of labour economics, development economics, health economics, public economics, etc. Researchers can thus adopt this framework to analyse the effects of those policies among individuals who are most responsive to the policies.


\clearpage 
\newpage 

{
\raggedright
\begingroup
\setlength{\bibsep}{0pt}
\setstretch{1}
\bibliography{ref.bib}

\begin{thebibliography}{}

\bibitem[\protect\citeauthoryear{Abadie and Imbens}{Abadie and Imbens}{2006}]{abadie2006matching}
Abadie, A. and G.~W. Imbens (2006).
\newblock Large sample properties of matching estimators for average treatment effects.
\newblock {\em Econometrica\/}~{\em 74\/}(1), 235--267.

\bibitem[\protect\citeauthoryear{Branstetter, Lima, Taylor, and Venâncio}{Branstetter et~al.}{2014}]{branstetter2014entry}
Branstetter, L., F.~Lima, L.~J. Taylor, and A.~Venâncio (2014).
\newblock Do entry regulations deter entrepreneurship and job creation? {E}vidence from recent reforms in {P}ortugal.
\newblock {\em The Economic Journal\/}~{\em 124\/}(577), 805--832.

\bibitem[\protect\citeauthoryear{Deng and Bijnens}{Deng and Bijnens}{2025}]{deng2025paper0}
Deng, H. and G.~Bijnens (2025).
\newblock A theory of new employer entry: Entrepreneurial decision under uncertainty.
\newblock {\em Working paper\/}.

\bibitem[\protect\citeauthoryear{Heckman, Ichimura, and Todd}{Heckman et~al.}{1997}]{heckman1997matching}
Heckman, J.~J., H.~Ichimura, and P.~E. Todd (1997, 10).
\newblock Matching as an econometric evaluation estimator: Evidence from evaluating a job training programme.
\newblock {\em The Review of Economic Studies\/}~{\em 64\/}(4), 605--654.

\bibitem[\protect\citeauthoryear{Heckman and Vytlacil}{Heckman and Vytlacil}{2001}]{heckman2001policy}
Heckman, J.~J. and E.~Vytlacil (2001).
\newblock Policy-relevant treatment effects.
\newblock {\em The American Economic Review\/}~{\em 91\/}(2), 107--111.

\bibitem[\protect\citeauthoryear{Heckman and Vytlacil}{Heckman and Vytlacil}{2005}]{heckman2005structural}
Heckman, J.~J. and E.~J. Vytlacil (2005).
\newblock Structural equations, treatment effects, and econometric policy evaluation.
\newblock {\em Econometrica\/}~{\em 73\/}(3), 669--738.

\bibitem[\protect\citeauthoryear{Hirano, Imbens, and Ridder}{Hirano et~al.}{2003}]{hirano2003efficient}
Hirano, K., G.~W. Imbens, and G.~Ridder (2003).
\newblock Efficient estimation of average treatment effects using the estimated propensity score.
\newblock {\em Econometrica\/}~{\em 71\/}(4), 1161--1189.

\bibitem[\protect\citeauthoryear{Hombert, Schoar, Sraer, and Thesmar}{Hombert et~al.}{2020}]{hombert2020france}
Hombert, J., A.~Schoar, D.~Sraer, and D.~Thesmar (2020).
\newblock Can unemployment insurance spur entrepreneurial activity? {E}vidence from {F}rance.
\newblock {\em The Journal of Finance\/}~{\em 75\/}(3), 1247--1285.

\bibitem[\protect\citeauthoryear{Imbens}{Imbens}{2004}]{imbens2004nonparametric}
Imbens, G.~W. (2004).
\newblock Nonparametric estimation of average treatment effects under exogeneity: A review.
\newblock {\em Review of Economics and Statistics\/}~{\em 86\/}(1), 4--29.

\bibitem[\protect\citeauthoryear{Imbens and Angrist}{Imbens and Angrist}{1994}]{imbens1994identification}
Imbens, G.~W. and J.~D. Angrist (1994).
\newblock Identification and estimation of local average treatment effects.
\newblock {\em Econometrica\/}~{\em 62\/}(2), 467--475.

\bibitem[\protect\citeauthoryear{Lee}{Lee}{2009}]{lee2009training}
Lee, D.~S. (2009).
\newblock Training, wages, and sample selection: Estimating sharp bounds on treatment effects.
\newblock {\em Review of Economic Studies\/}~{\em 76\/}(3), 1071--1102.

\bibitem[\protect\citeauthoryear{Rosenbaum and Rubin}{Rosenbaum and Rubin}{1983}]{rosenbaum1983central}
Rosenbaum, P.~R. and D.~B. Rubin (1983).
\newblock The central role of the propensity score in observational studies for causal effects.
\newblock {\em Biometrika\/}~{\em 70\/}(1), 41--55.

\bibitem[\protect\citeauthoryear{Sugiyama, Krauledat, and M{{\"u}}ller}{Sugiyama et~al.}{2007}]{sugiyama2007importance.weighted}
Sugiyama, M., M.~Krauledat, and K.-R. M{{\"u}}ller (2007).
\newblock Covariate shift adaptation by importance weighted cross validation.
\newblock {\em Journal of Machine Learning Research\/}~{\em 8\/}(35), 985--1005.

\end{thebibliography}
\endgroup
\pagebreak} 


\appendix 
\section*{Appendix} 

\section{Derivations} 

\subsection{IPW recovers PATE}
\label{app:ipw.recovers.pate}

The population average treatment effect (PATE) averages treatment effects over the entire distribution of firm types, regardless of whether they hire before or after the reform. In many observational settings, this object is estimated using inverse probability weighting (IPW), where treated and untreated observations are weighted by the inverse of their estimated treatment probabilities in order to simulate a balanced pseudo-population.

In the context of this paper, hiring decisions are affected by a policy reform. Let \( s = 0 \) denote the pre-reform regime and \( s = 1 \) the post-reform regime. In each regime \( s \), firms hire with probability \( p^s(x) = \mathbb{P}(D = 1 \mid X = x, s) \), where \( X \) denotes observed firm characteristics. Outcomes \( y^s \) are observed only for hiring firms, i.e., those with \( D = 1 \).

For each regime \( s \), one can define an inverse probability-weighted outcome mean as
\begin{equation}
\bar{y}^s_{\text{IPW}} = \mathbb{E}_F\left[ \frac{D^s \cdot y^s}{p^s(x)} \right],
\end{equation}
where the expectation is taken over the full population distribution of \( x \), and \( D^s \) is the observed treatment status in regime \( s \). Under standard assumptions—namely, that the potential outcomes \( y^s(x) \) are mean-independent of treatment conditional on \( X \) and regime—the inverse probability-weighted means identify the average potential outcomes in each regime:
\[
\bar{y}^s_{\text{IPW}} = \mathbb{E}_F[y^s(x)].
\]

It follows that the difference between the two regime-specific IPW means recovers the PATE:
\[
\tau^{\texttt{PATE}} = \mathbb{E}_F[y^1(x)] - \mathbb{E}_F[y^0(x)] = \bar{y}^1_{\text{IPW}} - \bar{y}^0_{\text{IPW}}.
\]

This derivation shows that even though hiring probabilities differ across regimes, and are denoted by \( p^0(x) \) and \( p^1(x) \), the use of inverse probability weights within each regime appropriately adjusts for selection into treatment and yields an estimate of the average treatment effect over the full population of firms.

\subsection{OMD decomposition} 
\subsubsection{Derivation of the decomposition} 
\label{app:omd-decomp}

This section formally derives the decomposition of the observed mean difference (OMD) in outcomes between post- and pre-reform employer cohorts. The OMD estimand can be expressed as the sum of three interpretable components: a treatment effect, a selection bias, and a reweighting effect.

The observed mean difference is defined as $\tau^{\texttt{OMD}} = \bar{y}^1 - \bar{y}^0$. Using the potential outcomes notation, this expression can be decomposed as:
\begin{align}
\tau^{\texttt{OMD}} 
&= \mathbb{E}_{Q^1}[y^1(\theta)] - \mathbb{E}_{Q^0}[y^0(\theta)] \nonumber \\
&= \underbrace{\mathbb{E}_{Q^1}[y^1(\theta) - y^0(\theta)]}_{\text{Post-reform-weighted treatment effect}} 
+ \underbrace{\mathbb{E}_{Q^1}[y^0(\theta)] - \mathbb{E}_{Q^0}[y^0(\theta)]}_{\text{Bias}}. 
\end{align}

The second term represents the change in untreated outcomes due to the change in the composition of firms that decide to hire. To further unpack this, observe that:
\begin{align}
\mathbb{E}_{Q^1}[y^0(\theta)] - \mathbb{E}_{Q^0}[y^0(\theta)] 
= \int y^0(\theta) \cdot f(\theta) \cdot \left( \frac{p^1(\theta)}{N^1} - \frac{p^0(\theta)}{N^0} \right) \, d\theta.
\end{align}

This integrand can be split into two terms:
\begin{align}
& \int y^0(\theta) \cdot f(\theta) \cdot \left( \frac{1}{N^1} - \frac{1}{N^0} \right) p^0(\theta) \, d\theta 
+ \int y^0(\theta) \cdot f(\theta) \cdot \frac{1}{N^1} \cdot \left( p^1(\theta) - p^0(\theta) \right) \, d\theta.
\end{align}

The first term is the \textbf{re-weighting effect bias}, arising purely from the change in sample weights even if firm composition had remained unchanged. The second term is the \textbf{selection bias}, driven by the fact that different types of firms entered the market after the reform.

Combining all components yields the full decomposition in Eq \eqref{eq:omd.de}, where only the first term identifies a meaningful causal effect; the latter two distort the interpretation of the OMD estimand unless properly accounted for.

\subsubsection{Alternative decomposition} 
\label{app:alt.omd}

If the identification goal is ATE under the pre-reform importance-weighted distribution, $\mathbb{E}_{Q^0}[\tau(\theta)]$, then the OMD estimand can be alternatively accordingly, 
\begin{equation}
    \tau^\texttt{OMD} = \underbrace{\mathbb{E}_{Q^0}[\tau(\theta)]}_\text{Pre-reform-weighted ATE}
    + \underbrace{\mathbb{E}_{Q^1}[y^1(\theta)]-\mathbb{E}_{Q^0}[y^1(\theta)]}_\text{Selection bias} 
    + \underbrace{\mathbb{E}_{F}\bigg[ y^1(\theta)p^1(\theta)\bigg(\frac{1}{N^0} - \frac{1}{N^1}\bigg) \bigg]}_\text{Re-weighting effect bias} . 
\end{equation}

\subsection{PSM recovers importance-weighted estimands}
\label{app:psm}

Let \( s \in \{0,1\} \) denote the regime (pre- or post-reform). Let \( D_i^s \in \{0,1\} \) indicate whether firm \( i \) hires in regime \( s \), and suppose we observe \( y_i^s = y_i^1 \) if \( D_i^s = 1 \), and \( y_i^s = y_i^0 \) if \( D_i^s = 0 \). Assume that potential outcomes satisfy selection-on-observables: for all \( s \), \( (y^0, y^1) \perp D^s \mid X \).

Define the PSM estimator 
\[
\hat{\tau}^{\texttt{PSM},s} = \frac{1}{n_1^s} \sum_{i: D_i^s = 1} \left( y_i^1 - y_{\mathcal{M}(i)}^0 \right),
\]
where \( y_i^1 \) is the outcome for a post-reform treated firm (hiring under regime \( s = 1 \)), \( y_{\mathcal{M}(i)}^0 \) is the outcome of the matched pre-reform treated firm with similar \( \hat{p}^s(x) \), and $n_1^s = \sum_i{\mathbb{1}[D_i^s=1]} $ denotes the number of treated units (i.e., hiring firms) observed under regime $s$.

Assume that the matching is exact or asymptotically consistent in \( x \), and that selection into hiring within each regime is independent of potential outcomes conditional on \( X \). Then, for each matched pair \( (i, \mathcal{M}(i)) \), we have:
\[
y_i^1 - y_{\mathcal{M}(i)}^0 \xrightarrow{p} \tau(x_i),
\]
where \( \tau(x) = \mathbb{E}[y^1(x) - y^0(x) \mid X = x] \).

To complete the argument, we now justify the convergence of the empirical average over treated firms to the \( Q^s \)-weighted expectation. Let \( \tau(x) \) be measurable and integrable, and suppose firms \( i \in \{1, \dots, n\} \) are independently drawn from a population with covariate distribution \( f(x) \), and regime-\( s \) treatment probability \( p^s(x) = \mathbb{P}(D_i^s = 1 \mid X_i = x) \). Then the treated sample \( \{ X_i : D_i^s = 1 \} \) is drawn from the selection-weighted distribution \( p^s(x) \cdot f(x) \), and by the law of large numbers, the sample average satisfies:

\[
\frac{1}{n_1^s} \sum_{i : D_i^s = 1} \tau(X_i) \xrightarrow{p} \frac{1}{N^s} \int \tau(x) \cdot p^s(x) \, dF(x) = \tau^{Q^s}. 
\]
That is, the empirical distribution of \( X \) among treated firms converges to the density \( q^s(x) = \frac{p^s(x)}{N^s} f(x) \), and thus the matching estimator over treated firms recovers the \( Q^s \)-weighted average treatment effect.

\subsection{Identifying the marginality-weighted estimand from observables}
\label{app:delta_p_observable}

To recover the marginality-weighted average treatment effect \( \tau^{\Delta p} \) using observable data, I impose the following assumptions:

\paragraph{Assumption 1 (Conditional independence).} Given \( X \), the distribution of \( \theta \) is independent of treatment assignment: \( \theta \perp D \mid X \). This is plausible if $X$ includes good proxies of firm productivity $\theta$. 

\paragraph{Assumption 2 (Measurability).} The functions \( \tau(\theta) \) and \( \Delta p(\theta) \) are measurable with respect to \( X \), so that we can define:
\[
\tau(x) = \mathbb{E}[\tau(\theta) \mid X = x], \quad \Delta p(x) = \mathbb{E}[\Delta p(\theta) \mid X = x].
\]

\paragraph{Assumption 3 (Support and positivity).} The marginal distribution \( f(x) \) is positive and has common support under both regimes. Moreover, \( \Delta p(x) > 0 \) on the support of \( X \).

\bigskip
\noindent

\paragraph{Proposition.} Under Assumptions 1–3, the marginality-weighted average treatment effect \( \tau^{\Delta p} \), defined as
\[
\tau^{\Delta p} = \frac{\int \tau(\theta) \cdot \Delta p(\theta) \, dF(\theta)}{\int \Delta p(\theta) \, dF(\theta)},
\]
is equal to the observable counterpart
\[
\tilde{\tau}^{\Delta p} = \frac{\int \tau(x) \cdot \Delta p(x) \, dF(x)}{\int \Delta p(x) \, dF(x)}.
\]

\paragraph{Proof.}
Using the law of iterated expectations, the numerator becomes:
\[
\int \tau(\theta) \cdot \Delta p(\theta) \, dF(\theta)
= \int \left( \int \tau(\theta) \cdot \Delta p(\theta) \, dF(\theta \mid X = x) \right) f(x) \, dx.
\]
By Assumption 1, \( \tau(\theta) \) and \( \Delta p(\theta) \) are independent conditional on \( X \), so:
\[
\mathbb{E}[\tau(\theta) \cdot \Delta p(\theta) \mid X = x] = \tau(x) \cdot \Delta p(x).
\]
Hence,
\[
\int \tau(\theta) \cdot \Delta p(\theta) \, dF(\theta) = \int \tau(x) \cdot \Delta p(x) \cdot f(x) \, dx.
\]

Similarly,
\[
\int \Delta p(\theta) \, dF(\theta) = \int \Delta p(x) \cdot f(x) \, dx.
\]

It follows that
\[
\tau^{\Delta p} = \tilde{\tau}^{\Delta p}.
\qquad
\]

\paragraph{Empirical implementation.} Given this identification, the estimand \( \tau^{\Delta p} \) can be consistently estimated by:
\begin{enumerate}
    \item Estimating \( \Delta p(x) = \mathbb{E}[D = 1 \mid X = x, s = 1] - \mathbb{E}[D = 1 \mid X = x, s = 0] \) via logistic regression or binned sample differences,
    \item Estimating \( \tau(x) = \mathbb{E}[y \mid D = 1, X = x, s = 1] - \mathbb{E}[y \mid D = 1, X = x, s = 0] \),
    \item Computing the weighted average:
    \[
    \hat{\tau}^{\Delta p} = \frac{\sum_{x} \hat{\tau}(x) \cdot \hat{\Delta p}(x) \cdot \hat{f}(x)}{\sum_{x} \hat{\Delta p}(x) \cdot \hat{f}(x)},
    \]
    where \( \hat{f}(x) \) is the empirical frequency of \( x \) in the pooled sample.
\end{enumerate}

\paragraph{Relation to existing literature.} This approach closely relates to the MTE framework \citep{heckman2005structural}, which emphasizes the importance of accounting for heterogeneity in treatment response and selection, and the policy-relevant treatment effect (PRTE) \citep{heckman2001policy}, in which average treatment effects are weighted by the shift in the treatment propensity due to policy. However, my framework avoids instrument-based identification and focuses instead on decomposing observable variation in hiring probabilities induced by a sharp policy discontinuity.

\section{Poster} 
This section provides a poster version visualising the note's idea.\footnote{The poster uses Canva template `Planet Research Digital Education Poster Lined Graphic Style' by Amanda Kauffroath.} 

\begin{center}
    \includegraphics[width=0.8\linewidth]{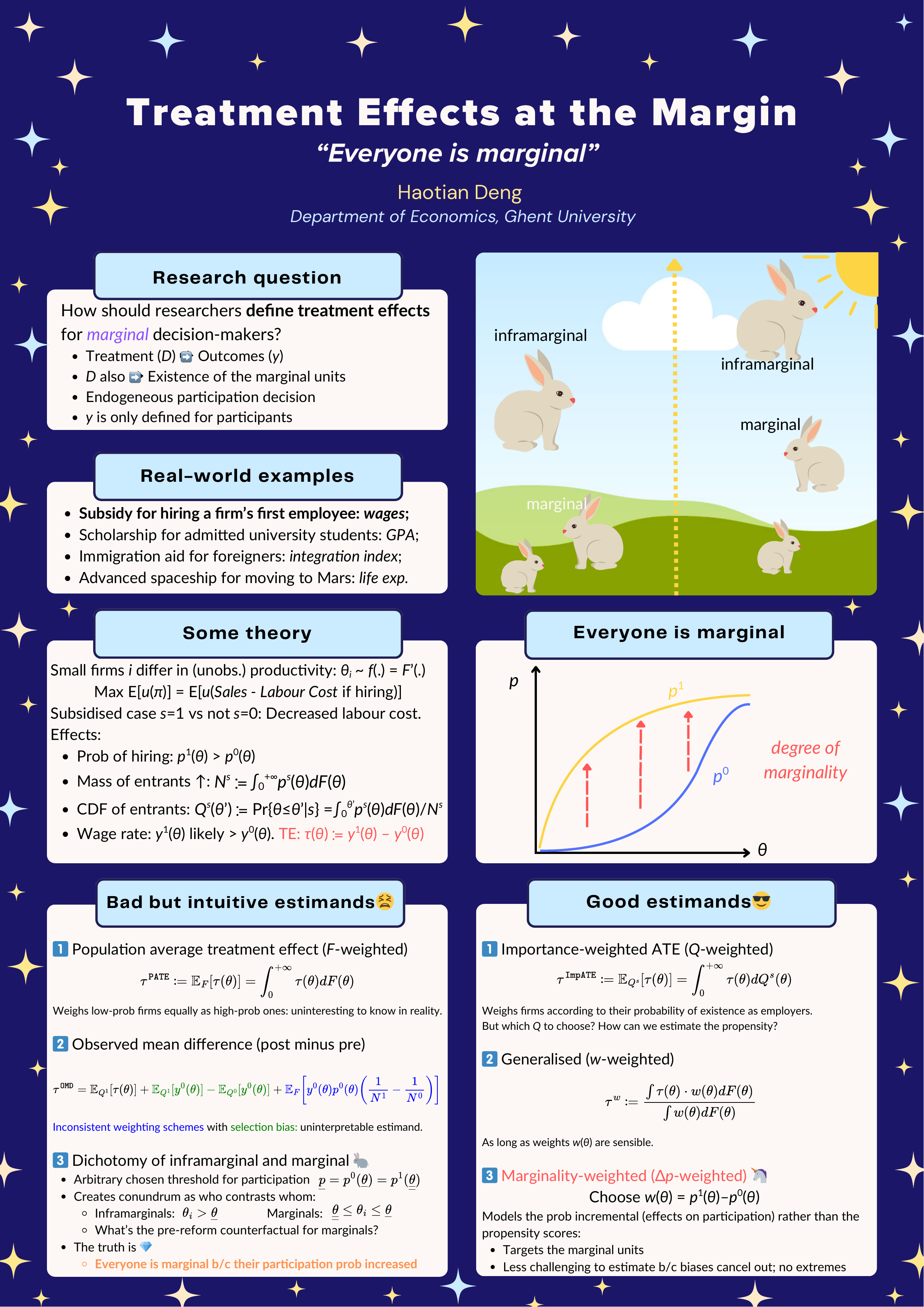}
\end{center}

\end{document}